\documentclass[10pt,final,journal,twocolumn]{IEEEtran}

\ifCLASSOPTIONcaptionsoff
  \newpage
\fi

\usepackage{cite}
\usepackage[cmex10]{amsmath}
\usepackage{amsmath}
\usepackage{algorithm}
\usepackage{algorithmic}
\usepackage{stfloats}
\usepackage{multicol,multienum}
\usepackage{multirow}
\usepackage[dvips]{graphicx}
\usepackage{wrapfig}
\usepackage{color}
\usepackage{subfigure}
\begin{document}
%\title{A Dynamic User-Scheduling-based Hierarchical Power Control in Two-tier Femtocell Networks from the Perspective of Energy Efficiency}
\title{A Collaborative Multi-agent Reinforcement Learning Anti-jamming Algorithm in
Wireless Networks}

\author{
       Fuqiang~Yao,
        Luliang~Jia
        % <-this % stops a space

%\thanks{This work was supported in part by the Natural Science Foundation for Distinguished Young Scholars of Jiangsu Province under Grant BK20160034, in part by the National Science Foundation of China under Grant 61631020, Grant 61671473, Grant 61401508, and Grant 61401505, in part by Jiangsu Provincial Natural Science Foundation of China Grant BK20130069, and Grant BK20151450, and in part by the Open Research Foundation of Science and Technology in Communication Networks Laboratory. Part of this paper had been submitted to the 2017 IEEE International Conference on Communication Technology (ICCT) (Corresponding author: Y. Sun.)}

\thanks{
F.~Yao is with the Sixty-third/63rd Research Institute, National University of Defense Technology, Nanjing 210007, China (e-mail:yfq2030@163.com).}
%\thanks{
%F.~Yao is with  Nanjing Telecommunication
%Technology Institute, Nanjing 210007, China (e-mail: yfq2030@163.com).}
%\thanks{Y.~Sun is with National Digital Switching System Engineering $\&$ Technological Research Center, Zhengzhou 450001, China (e-mail: sunyouming10@163.com).}
%\thanks{S.~Feng is with the Cognitive Systems Laboratory, McMaster University, Hamilton, ON L8S 4L8, Canada (e-mail: fengs13@mcmaster.ca).}
\thanks{
L.~Jia is with the College of Communications Engineering, Army Engineering University of PLA, Nanjing 210007, China. (e-mail:
jiallts@163.com).(Corresponding author: Luliang Jia.)}
% <-this % stops a space
%\thanks{
%F.~Yao , Y. Niu, and Y. Zhu are with  Nanjing Telecommunication
%Technology Institute, Nanjing 210007, China (e-mail:yfq2030@163.com; niuyingtao78@hotmail.com; zhumaka1982@163.com).}% <-this % stops a space
}

\maketitle

\begin{abstract}
In this letter, we investigate the anti-jamming defense problem in multi-user scenarios, where the coordination among users is taken into consideration. The Markov game framework is employed to model and analyze the anti-jamming defense problem, and a collaborative multi-agent anti-jamming algorithm (CMAA) is proposed to obtain the optimal anti-jamming strategy. In sweep jamming scenarios, on the one hand, the proposed CMAA can tackle the external malicious jamming. On the other hand, it can effectively cope with the mutual interference among users. Simulation results show that the proposed CMAA is superior to both sensing based method and independent Q-learning method, and has the highest normalized rate.
\end{abstract}


% \fi
\IEEEpeerreviewmaketitle

\section{Introduction}
Jamming attack is a serious threat in wireless networks, and various anti-jamming methods have been developed in recent years \cite{existwork1}-\cite{existworkadd}. Due to factors of the jammers' activities, the quality of channels varies between ``good"  and ``poor" dynamically. The Markov decision process (MDP) \cite{existwork8} is a suitable paradigm to model and analyze the anti-jamming defense problem. Unfortunately, it is difficult to obtain the state transition probability function in an adversarial environment. In these scenarios, reinforcement learning (RL) techniques are available, such as the classic Q-learning method \cite{existwork9}. Based on the Q-learning method, the anti-jamming decision-making problem in single-user scenarios were investigated in \cite{existwork10}-\cite{existwork12}. Then, the authors in \cite{existwork13}-\cite{existwork15} extended it to the multi-user scenarios, and they resorted to the Markov game framework \cite{existwork16}, which is the extension of the Markov decision process and can characterize the relationship among multiple users. Moreover, the corresponding multi-user reinforcement learning anti-jamming algorithm was designed. However, each user employed a standard Q-learning method in \cite{existwork13}-\cite{existwork15}, and the coordination among users was not considered.

In order to achieve better anti-jamming performance, the coordination among users is necessary. Through collaborative learning, on the one hand, it can tackle the external malicious jamming, and on the other hand, it can effectively cope with the mutual interference caused by competition among users. In this letter, a collaborative anti-jamming framework is formulated, in which the ``coordination" and ``competition" are simultaneously considered. To model and analyze the anti-jamming defense problem, the Markov game framework is adopted, and a collaborative multi-agent reinforcement learning anti-jamming algorithm is proposed. The main contributions of this letter are given as follows:
\begin{itemize}
\item  Based on the Markov game, the anti-jamming defense problem is investigated in multi-user scenarios, and the coordination among users is considered.
\item We develop a collaborative multi-agent reinforcement learning anti-jamming algorithm to obtain the optimal anti-jamming strategy.
\end{itemize}

% To the best of our knowledge,  there are few researches on dynamically hierarchical joint user-scheduling and power control in tiered femtocell networks considering energy consumption.

\section{System Model and Problem Formulation}
\subsection{System Model}
%%%%%%%%%%%%%%%%%%%%%%%%%%%%%%%%%%%%%%%%%%%%%%%%%%%%%%
\begin{figure}[!t]
\centering
\includegraphics[width=3.5in]{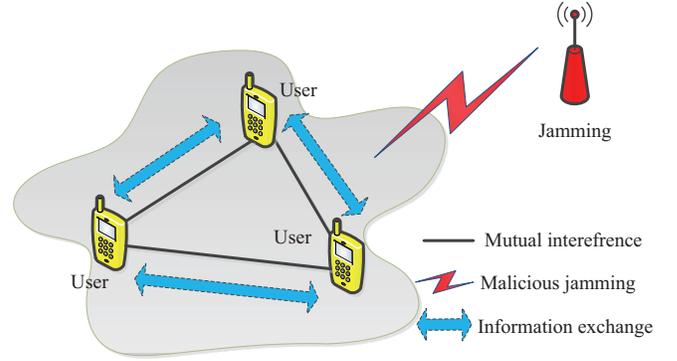}
\caption{System model.}
\label{Fig1}
\end{figure}

As illustrated in Fig. \ref{Fig1}, there are \emph{N} users and one jammer in the considered model. The user set is denoted as ${\cal N} = \left\{ {1, \cdots ,N} \right\}$, and the available channel set is defined as ${\cal M} = \left\{ {1, \cdots ,M} \right\}$. The number of available channels is \emph{M} ($N < M$). The coordination among users can be achieved through information exchange. It is noted that the jamming pattern is the sweep jamming, and one channel is jammed at each time slot. The jamming channel set is represented as ${\cal C} = \left\{ {1, \cdots ,C} \right\}$. In this letter, we assume that the available channel set is the same as the jamming channel set. If two or more users select the same channel, it will lead to the mutual interference. In order to realize the reliable transmission, it is necessary to simultaneously consider the external malicious jamming and mutual interference due to competition among users. In this letter, the mutual interference refers to the co-channel interference among users, and the strategy is the selection of available channels.

%%%%%%%%%%%%%%%%%%%%%%%%%%%%%%%%%%%%%%%%%%%%%%%%%%%%%%%
\subsection{Problem formulation}

The anti-jamming defense problem can be formulated as a Markov game \cite{existwork16}, which is the extension of the Markov decision process in multi-user scenarios. Mathematically, it can be expressed as ${\cal G} = \left\{ {{\cal S},{{\cal A}_1}, \cdots ,{{\cal A}_N},f,{r_1}, \cdots ,{r_N}} \right\}$, where ${\cal S}$ denotes the set of states, ${{\cal A}_n},n = 1, \cdots ,N$ is the set of the strategies, $f$ represents the state transition model, and ${r_n},n = 1, \cdots ,N$ is the reward. In this letter, referring to \cite{existwork10}, \cite{existwork11}, the state can be defined as $s = \{ {\bf{a}},{f_{jx}}\}$, where ${\bf{a}} = \left( {{a_1},{a_2}, \cdots ,{a_N}} \right)$ represents a joint action profile, and the set of the joint action profiles is ${\cal A}{\rm{ = }} \otimes {{\cal A}_n},n = 1, \cdots ,N$, where $\otimes$ represents the Cartesian product. Similar to \cite{existwork17}, the global reward can be defined as:

\begin{equation}
\label{eq1}
R = \sum\nolimits_{n = 1}^N {{r_n}(s,{\bf{a}})} ,
\end{equation}
where $s \in {\cal S}$ denotes a state. It is assumed that the jamming channel is denoted as ${f_{jx}}$, the selected channel of user \emph{n} is represented as ${f_{n,x}}$, and the reward of user \emph{n} at time slot \emph{t} can be expressed as:

\begin{equation}
{r_n}(s,{\bf{a}},t) = \left\{ \begin{array}{l}
1,\quad \text{if}\;{f_{n,x}} \ne {f_{jx}}\;\& {f_{n,x}} \ne {f_{m,x}}\left( {m \in {{\cal N} \mathord{\left/
 {\vphantom {{\cal N} n}} \right.
 \kern-\nulldelimiterspace} n}} \right),\\
0,\quad \text{otherwise}.
\end{array} \right.
\end{equation}

\section{Collaborative Multi-agent Anti-jamming Algorithm}

\begin{figure}[!t]
\centering
\includegraphics[width=3.5in]{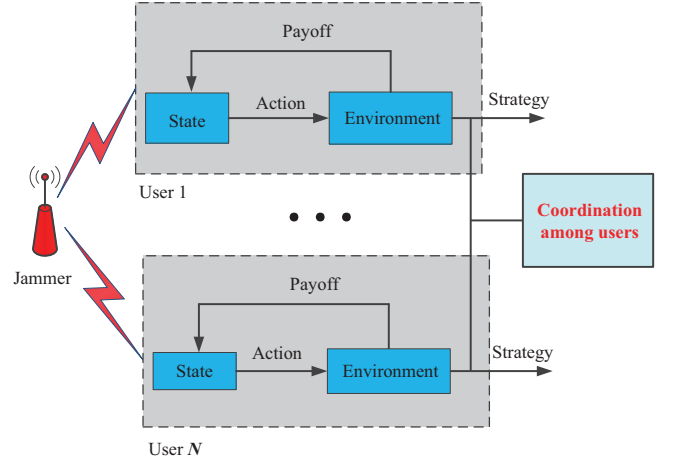}
\caption{Illustration of the collaborative anti-jamming framework.}
\label{Fig2}
\end{figure}

In this letter, we consider the two characteristics ``coordination" and ``competition" among users simultaneously. A collaborative anti-jamming framework is shown in Fig. \ref{Fig2}. In wireless network, the coordination has various meanings, such as relay and information exchange. Here, the coordination is realized by information exchange among users. Based on the coordination among users, the method of ``decision-feedback-adjustment" is applied to obtain the optimal anti-jamming strategy.

To solve the formulated anti-jamming Markov game, a multi-agent Q-learning algorithm is proposed. Similar to \cite{existwork17}, user \emph{n} updates its Q values according to the following rules:

\begin{equation}
\label{eq3}
{Q_n}\left( {s,{\bf{a}}} \right) = (1 - \lambda ){Q_n}\left( {s,{\bf{a}}} \right) + \lambda \left[ {{r_n} + \gamma {V_n}(s')} \right],
\end{equation}

\begin{equation}
\label{eq4}
{V_n}(s') = {Q_n}\left( {s',{{\bf{a}}^{\rm{*}}}} \right), \text{where}\;{{\bf{a}}^{\rm{*}}} \in \mathop {\arg \max }\limits_{{\bf{a}}'} \sum\limits_{n = 1}^N {{Q_n}\left( {s',{\bf{a}}'} \right)} ,
\end{equation}
where $\lambda$ is the learning rate. It is noted that the multi-agent Q-learning algorithm in (3) is decentralized, and each user updates its Q values separately. However, for the problem in (4), it is necessary to solve a global coordination game, which has common payoff \cite{existwork17}:

\begin{equation}
\label{eq5}
\;Q\left( {s,{\bf{a}}} \right) = \sum\limits_{n = 1}^N {{Q_n}\left( {s,{\bf{a}}} \right)} .
\end{equation}

\begin{figure}[tb]
\rule{\linewidth}{1pt}
\emph{\textbf{Algorithm 1}: Collaborative Multi-agent Anti-jamming Algorithm (CMAA)}\\
\rule{\linewidth}{1pt}
\begin{algorithmic}
\STATE \textbf{Initiate:}  ${{\cal S}}$, ${Q_n},\;n \in N$.
\\
\STATE \textbf{Loop:}  $t = 0, \cdots ,T$
\\
\STATE \quad \quad Each user observes its current state $s(t) = \{ {\bf{a}}(t),{f_{jx}}(t)\}$, and selects a channel according to the following rules:
\\
\STATE \quad \quad $\bullet$ User \emph{n} randomly chooses a channel profile ${\bf{a}} \in {\cal A}$ with probability $\varepsilon$;
\\
\STATE \quad \quad $\bullet$ User \emph{n} chooses a channel profile ${{\bf{a}}^{\rm{*}}} \in argmax\sum\nolimits_{n = 1}^N {{Q_n}\left( {s',{\bf{a}}'} \right)}$ with probability $1 - \varepsilon$.
\\
\STATE \quad \quad Each user measures its payoff ${r_n}(s,{\bf{a}})$.
\\
\STATE \quad \quad The state is transferred into $s(t + 1) = \{ {\bf{a}}(t + 1),{f_{jx}}(t + 1)\}$, and the Q values are updated according to the rules in (3).
\\
\STATE \textbf{End loop}
\\
\end{algorithmic}
\rule{\linewidth}{1pt}
\end{figure}

Each user broadcasts its current Q value to other users. The exploration rate $\varepsilon  \in (0,1)$ is introduced to avoid falling into a local optimum. Users randomly select a joint action ${\bf{a}} \in {\cal A}$ with probability $\varepsilon$, and users select the joint action ${{\bf{a}}^{\rm{*}}} \in argmax\sum\nolimits_{n = 1}^N {{Q_n}\left( {s',{\bf{a}}'} \right)}$ with probability $1 - \varepsilon$. Based on the above analysis, a collaborative multi-agent anti-jamming algorithm (CMAA) is proposed, and its implementation procedure is shown in Algorithm 1.

\begin{figure}[!t]
\centering
\includegraphics[width=3.5in]{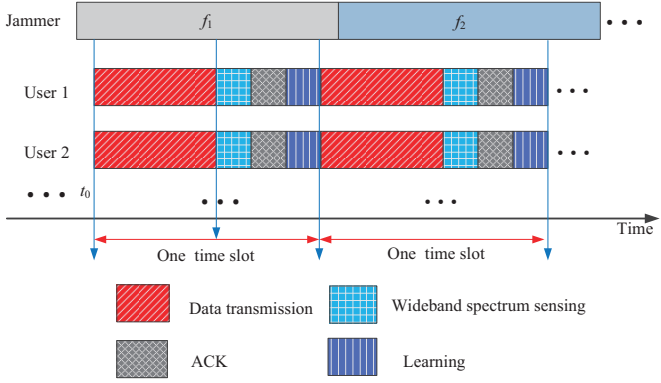}
\caption{Illustration of the transmission slot structure.}
\label{Fig3}
\end{figure}

Similar to \cite{existwork10}, the wideband spectrum sensing is adopted to sense the jammer's activities, and all Q values are updated simultaneously. A transmission slot structure diagram is presented in Fig. \ref{Fig3}. At the end of current slot, each user obtains a reward, and updates its strategy according to the received reward.

 \section{Numerical Results and Discussions}
%%%%%%%%%%%%%%%%%%%%%%%%
In this subsection, we present some simulation results. A system with two users and one jammer is considered, in which five channels are available. ${t_{Rx}}$, ${t_{WBSS}}$, ${t_{ACK}}$, and ${t_{Learning}}$ denote the transmission time, wideband sensing time, ACK time, and learning time, respectively. The jammer begins to jam the transmission at time slot $t = 0.2ms$. Referring to \cite{existwork11}, the simulation parameters are given as: $\lambda  = 0.8$, $\gamma  = 0.6$, $\varepsilon  = 0.2$, ${t_{Rx}} = 0.98ms$, ${t_{WBSS}}{\rm{ + }}{t_{ACK}}{\rm{ + }}{t_{Learning}} = 0.2ms$. Moreover, the dwelling time of the sweeping jammer on each channel is ${t_{dwell}} = 2.28ms$, the number of time slots for simulations is $K = 10000$, and the simulation time is $T = K * ({t_{Rx}} + {t_{WBSS}}{\rm{ + }}{t_{ACK}}{\rm{ + }}{t_{Learning}})$.

\begin{figure}[!t]
\centering
\includegraphics[width=3.5in]{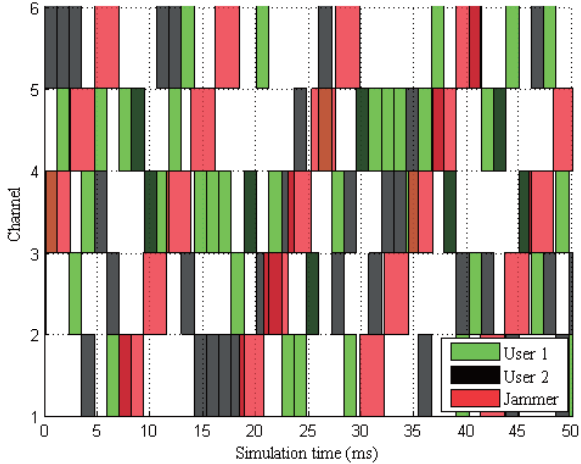}
\caption{Time frequency information at initial state.}
\label{Fig4}
\end{figure}

\begin{figure}[!t]
\centering
\includegraphics[width=3.5in]{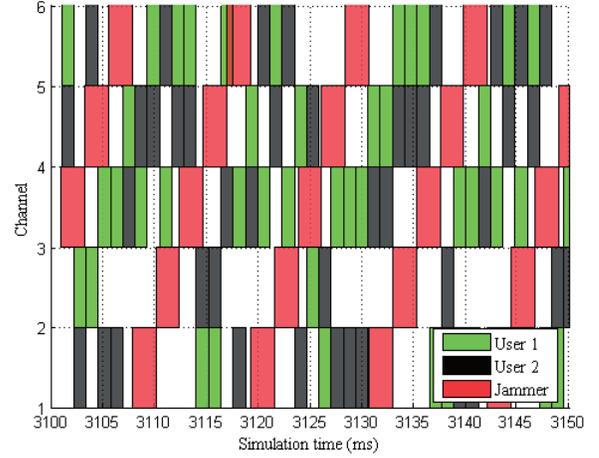}
\caption{Time-frequency information at convergent state.}
\label{Fig5}
\end{figure}

Fig. \ref{Fig4} and Fig. \ref{Fig5} respectively show the time-frequency information at the initial and convergent state. As indicated in Fig. \ref{Fig4}, at the initial stage, the users employ random actions, and the signals of users and jammer are overlapped. Moreover, the signals among users are also overlapped. Fig. \ref{Fig5} shows the time-frequency information of the proposed CMAA at convergent stage. As can be seen from Fig. \ref{Fig5}, at convergent stage, the signals of users can avoid the signal of the jammer. Meanwhile, the signals among users can effectively cope with the mutual interference, and the actions of users are coordinated.

\begin{figure}[!t]
\centering
\includegraphics[width=3.5in]{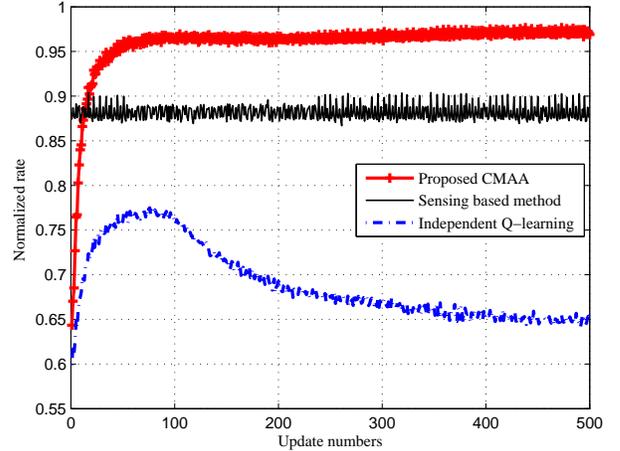}
\caption{Performance comparison of the normalized rate.}
\label{Fig6}
\end{figure}

To validate the proposed CMAA, we compare it with the following two methods:
\begin{itemize}
\item  Sensing based method: In this method, the users cannot learn the actions of the jammer, and channels are selected based on the sensing results. Moreover, we resort to a coordination approach, as in \cite{existwork17}, in which user \emph{n} ($n > 1$) selects its channel $a_n^*$ until the previous users $1, \cdots ,n - 1$ broadcast their chosen channels in the ordering. Then, user \emph{n} broadcasts its channel.
\item Independent Q-earning \cite{existwork13}: Each user adopts a standard Q-learning method. The coordination among users is not considered, and other users are treated as part of its environment.
\end{itemize}

In this section, the normalized rate is introduced to validate the performance of the proposed CMAA, and it can be defined as $\rho  = {{P{K_{succ}}} \mathord{\left/
 {\vphantom {{P{K_{succ}}} {P{N_0}}}} \right.
 \kern-\nulldelimiterspace} {P{N_0}}}$, where $P{K_{succ}}$ represents the number of packets for successful transmission, and $P{N_0}$ denotes the length of packet statistics, which means that the normalized rate $\rho$ is calculated after $P{N_0}$ packets are transmitted. In this simulation, we have $P{N_0} = 20$. Then, the following results are obtained by making 200 independent runs and then taking the mean.

Fig. \ref{Fig6} shows the performance of the normalized rate, it can be seen that the proposed CMAA is superior to both sensing based method and independent Q-learning method. Moreover, the proposed CMAA has the highest normalized rate $\rho$. The reason is that the sensing based method cannot learn the actions of the jammer, and channels are chosen based on the sensing results. Meanwhile, the independent Q-learning method does not consider the coordination among users, and each user chooses its channel independently. For the proposed CMAA, it can not only learn the actions of the jammer, but also consider the coordination among users.

\section{Conclusion}
In this letter, we consider the ``coordination" and ``competition" simultaneously, and the Markov game framework is employed to model and analyze the anti-jamming defense problem. Then, a collaborative multi-agent anti-jamming algorithm (CMAA) is proposed to obtain the optimal anti-jamming strategy. Through collaborative learning, it can cope with the external malicious jamming and the mutual interference caused by competition among users simultaneously. To validate the effectiveness of the proposed CMAA, simulation results are presented. Compared with the sensing based method and independent Q-learning method, the proposed CMAA has the highest normalized rate.


\begin{thebibliography}{1}
%%%%%%%%%%%%%%%%%%%%%%%%%%%%%%%%%%%%%%%%%%%%%自己参考文献%%%%%%%%%%%%%%%%%%%%%%%%%%%%%%%%%%%%%%%%%%%%%%%%%%%%%%%%%%%%%%%%%%%%%%%%%%%%%%%
%%%%%%%%%%%%%%%%%%%%%%%%%%%%%%%%%%%%%%%%%%%%%%%%%%%%%%%%%%%%%%%%%%%%%%
%\bibitem{existwork1}
%Y. Zou, J. Zhu, X. Wang, and L. Hanzo, ``A survey on wireless security: technical challenges, recent advances, and future trends," \emph{Proceedings of the IEEE}, vol. 104, no. 9, pp. 1727-1765, Sep. 2016.


\bibitem{existwork1}
K. Grover, A. Lim, and Q. Yang,``Jamming and anti-jamming techniques in wireless networks: A survey," \emph{Int. J. Ad Hoc and Ubiquitous Comput.}, vol. 17, no. 4, pp. 197-215, Dec. 2014.

\bibitem{existwork2}
L. Jia, Y. Xu, Y. Sun, S. Feng, and A. Anpalagan,``Stackelberg game approaches for anti-jamming defence in wireless networks," arXiv preprint arXiv: 1805.12308, 2018. (to be published in IEEE Wireless Communications)


\bibitem{existwork3}
D. Yang, G. Xue, J. Zhang, A. Richa, and X. Fang,``Coping with a smart jammer in wireless networks: A stackelberg game approach," \emph{IEEE Trans. Wireless Commun.}, vol. 12, no. 8, pp. 4038-4047, Aug. 2013.

\bibitem{existwork4}
L. Xiao, T. Chen, J. Liu, and H. Dai,``Anti-jamming transmission stackelberg game with observation errors," \emph{IEEE Commun. Lett.}, vol. 19, no. 6, pp. 949-952, Jun. 2015.

\bibitem{existwork5}
L. Jia, F. Yao, Y. Sun, Y. Niu, and Y. Zhu,``Bayesian Stackelberg game for anti-jamming with incomplete information," \emph{IEEE Commun. Lett.}, vol. 20, no. 10, pp. 1991-1994, Oct. 2016.

\bibitem{existwork6}
L. Jia, F. Yao, Y. Sun, Y. Xu, S. Feng, and A. Anpalagan,``A hierarchical learning solution for anti-jamming Stackelberg game with discrete power strategies," \emph{IEEE Wireless Commun. Lett.}, vol. 6, no. 6, pp. 818-821, Dec. 2017.

\bibitem{existwork7}
F. Yao, L. Jia, Y. Sun, Y. Xu, S. Feng, and Y. Zhu,``A hierarchical learning approach to anti-jamming channel selection strategies,"  \emph{Wireless Netw.}, DOI: 10.1007/s11276-017-1551-9, to be published.


\bibitem{existworkadd}
L. Jia, Y. Xu, Y. Sun, S. Feng, L. Yu, and A. Anpalagan,``A multi-domain anti-jamming defence scheme in heterogeneous wireless networks,"  \emph{IEEE Access}, DOI: 10.1109/ACCESS.2018.2850879, to be published.


\bibitem{existwork8}
Q. Hu, and W. Yue, \emph{Markov decision processes with their applications.} Springer US,
2007.


\bibitem{existwork9}
C. J. C. H. Watkins, and P. Dayan,``Q-learning," \emph{Mach. Learn.}, vol. 8, pp. 279-292, 1992.


\bibitem{existwork10}
F. Slimeni, B. Scheers, Z. Chtourou, \emph{et al.},``Jamming mitigation in cognitive radio networks using a modified Q-learning algorithm," \emph{in Proc. International Conference on Military Comunications and Information Systems (ICMCIS) 2015}, pp. 1-7.


\bibitem{existwork11}
F. Slimeni, Z. Chtourou, B. Scheers, \emph{et al}.,``Cooperative Q-learning based channel selection for cognitive radio networks,"  \emph{Wireless Netw.}, DOI:10.1007/s11276-018-1737-9, to be published.


\bibitem{existwork12}
S. Machuzak, and S. K. Jayaweera,``Reinforcement learning based anti-jamming with wideband autonomous cognitive radios," \emph{in Proc. IEEE International Conference on Communications in China (ICCC) 2016}, pp. 1-5.


\bibitem{existwork13}
M. A. Aref, S. K. Jayaweera, and S. Machuzak,``Multi-agent reinforcement learning based cognitive anti-jamming," \emph{in Proc. IEEE Wireless Communications and Networking Conference (WCNC) 2017}, pp. 1-6.

\bibitem{existwork14}
M. A. Aref, and S. K. Jayaweera,``A novel cognitive anti-jamming stochastic game," \emph{in Proc. Cognitive Communications for Aerospace Applications Workshop (CCAA) 2017}, pp. 1-4.


\bibitem{existwork15}
M. A. Aref, and S. K. Jayaweera,``A cognitive anti-jamming and interference-avoidance stochastic game," \emph{in Proc. IEEE International Conference on Cognitive Informatics and Cognitive Computing (ICCI*CC) 2017}, pp. 520-527.


\bibitem{existwork16}
L. Busoniu, R. Babuska, and B. D. Schutter,``A comprehensive survey of multiagent reinforcement learning," \emph{IEEE Trans. Systems, Man, and Cybernetics}, vol. 38, no. 2, pp. 156-172, Mar. 2008.


\bibitem{existwork17}
N. Vlassis, \emph{A concise introduction to multiagent systems and distributed artificial intelligence.} Morgan and Claypool Publishers,
2007.


%\bibitem{existwork15}
%Y. Sun, Q. Wu, \emph{et al}., ``Distributed channel access for device-to-device communications: A hypergraph-based learning solution," \emph{IEEE Commun. Lett.}, vol. 21, no. 1, pp. 180-183, Jan. 2017.



%\bibitem{existwork15}
%L. Jia, Y. Xu, \emph{et al.},``A Distributed Anti-jamming Channel Selection Algorithm for Interference Mitigation-based Wireless Networks," submitted to ICCT 2017.




%%%%%%%%%%%%%%%%%%%%%%%%%%%%%%%%%%%%%%%%%%%%%%%%%%%%%%%%%%%%%%%%%%%%%%%%%%%%%%%%%%%%%%%%%%%%%%%%%%%%%%%%%%%%%%%%%%%%%%%%%%%%%%

\end{thebibliography}
\end{document}